\documentclass[dvips]{acta}
\usepackage{supertabular,lscape,epsfig}
\usepackage{amssymb}
\usepackage{amsmath}
\SetPages{1}{16}
\SetVol{60}{2010} 

\def \PASJ {{\it Publ.Astr.Soc.Japan}\/} 

\begin{document}

\begin{Titlepage}

\Title { Superhumps and their Amplitudes }

\Author {J.~~S m a k}
{N. Copernicus Astronomical Center, Polish Academy of Sciences,\\
Bartycka 18, 00-716 Warsaw, Poland\\
e-mail: jis@camk.edu.pl }

\Received{  }

\end{Titlepage}

\Abstract { 
Superhump amplitudes observed in dwarf novae during their superoutbursts 
depend on orbital inclination: the maximum amplitudes 
in systems with low inclinations are $A_\circ \approx 0.25$ mag., while 
at higher inclinations they increase from $A_\circ \sim 0.3$ to 
$A_\circ \sim 0.6$ mag. 

The mean maximum superhump amplitudes normalized to the average luminosity 
of the disk are: $<A_n>=0.34\pm 0.02$ in low inclination systems and only  
$<A_n>=0.17\pm 0.01$ in high inclination systems. This shows that 
at high inclinations the superhump lIght source is {\it partly} obscured 
by the disk edge and implies that it is located close to the disk surface 
but extends sufficiently high above that surface to avoid full obscuration. 
Superhump amplitudes in high inclination systems show modulation with beat 
phase ($\phi_b$), interpreted as being due to azimuth-dependent 
obscuration effects in a non-axisymmetric disk. 
In addition they show modulation with $2\phi_b$ which implies that the orientation 
of the superhump light source is correlated with the direction of the stream. 

The dependence of superhump amplitudes on orbital inclination and their modulation 
with beat phase eliminate the tidal-resonance model for superhumps. 
Instead they support an alternative interpretation of superhumps as being due to 
periodically modulated dissipation of the kinetic energy of the stream. 

Superhump amplitudes in permanent superhumpers are $<A>=0.12$, i.e. much 
smaller than the maximum amplitudes observed during superoutbursts. 
} 
{accretion, accretion disks -- binaries: cataclysmic variables, stars: dwarf novae }

\section {Introduction }

Superhumps were discovered by Vogt (1974) and Warner (1975) during the December 1972 
superoutburst of VW Hyi. It is now well established that they are present  
in all dwarf novae of the SU UMa type during their superoutbursts 
(see Warner 1995, Hellier 2001). 
Their periods are slightly longer than the orbital periods and usually show 
complex variations (cf. Kato et al. 2009). 
Superhumps are also present in the so-called permanent superhumpers -- 
the nova-like cataclysmic binaries with stationary accretion (cf. Patterson 1999) 
as well as in some dwarf novae at quiescence (Still et al. 2010 and references 
therein). 

Any periodic phenomenon is characterized by two paremeters: 
the period and the amplitude. 
In the case of superhumps, however, most authors concentrate primarily or even 
exclusively on the superhump periods, treating their amplitudes as less important. 
Of course, there are papers (e.g. Warner and O'Donoghue 1988, Olech et al. 2003) 
which present full description of superhumps, including their individual 
light curves, but -- regretfully -- they belong to minority. 
A typical paper in this field (see, for example, Kato et al. 2009) contains  
"journals of observations" (which are meaningless), tables containing 
moments of superhump maxima (but not their amplitudes!), figures showing the (O-C) 
diagrams and -- not always -- only the {\it mean} superhump light curves without 
any information on their variations. 

As a result of this attitude our knowledge about superhump amplitudes has been, 
until now, limited to the few following conclusions: 

{\parskip=0truept {
(1) Superhumps first appear around superoutburst maximum (or shortly earlier; 
see Semeniuk 1980) and reach their highest amplitude either at maximum or 
1-2 days later. 

(2) Superhump amplitudes decrease during superoutburst; this effect, however, 
has been well documented only for relatively few cases 
(e.g. Warner and O'Donoghue 1988, Fig.5; Patterson et al. 2000b, Fig.10; 
Rutkowski et al. 2007, Fig.4). 

(3) According to Warner (1995, p.194): "At their full development the superhumps 
have a range of 0.3-0.4 mag. and are equally prominent in all SU UMa stars 
{\it independent of inclination}".  

(4) According to Warner (1985, p.372): "There is no strong modulation 
of the superhump profile at the beat period." 

As it will turn out (see Section 3) the last two statements are not true. 
\parskip=12truept 


The aim of the present paper is to improve the situation by presenting new 
results concerning superhump amplitudes based on representative samples of dwarf 
novae at their superoutbursts and of permanent superhumpers.

\section {Definitions and Formulae }

\subsection {The Amplitudes of Superhumps }

In what follows we will discuss the {\it full} amplitudes of superhumps. 
With this definition the superhump amplitude -- in intensity units -- 
is given by  

\beq
a~=~{ { \ell_{max}~-~\ell_{min} } \over {\ell_{min}} }~
  =~{ {\ell_{sh}}\over {\ell_d} }~,
\eeq

\noindent
where $\ell_d=\ell_{min}$ is the luminosity of the disk, and $\ell_{sh}$ is the 
luminosity of the superhump at its maximum. It is more common, 
however, to express superhump amplitudes in magnitudes: 

\beq
A({\rm mag})~=~2.5~\log~(1~+~a)~. 
\eeq

\noindent
Note that in both cases the superhump amplitude is defined with respect to 
the luminosity of the disk.

\subsection {The Normalized Superhump Amplitudes }

The observed luminosity of the disk depends on inclination. At low inclinations 
this dependence can be described with a simple formula applicable to a flat disk 
with limb darkening $u=0.6$: 

\beq
\ell_d(i)~=~<\ell_d>~(1~+1.5~\cos i)~\cos i~. 
\eeq

\noindent
At inclinations higher than $i\sim 70^\circ$, however, it is necessary to include 
contributions from the disk edge and -- at $i>80^\circ$ -- effects of self-obscuration. 
For the purpose of the present discussion a sequence of steady-state disk models 
was calculated, using model parameters of Z Cha and $\dot M= 3\times 10^{17}$ g/s. 
Results can be be represented by an approximate formula replacing Eq.(3) for 
$i>60^\circ$: 

\beq
\ell_d(i)~=~<\ell_d>~(~3.414~-~0.05374~i~+~0.00020~i^{~2}~)~. 
\eeq

The superhump amplitude normalized to $<\ell_d>$ can then be computed from 

\beq
A_n({\rm mag})~=~2.5~\log~\left [~1~+~\Bigl(~10^{~0.4A}~-~1~\Bigr)~
           {{\ell_d(i)}\over {<\ell_d>}}~\right ]~.
\eeq

\subsection {The Beat Phases } 

The beat phase is related to the orbital and superhump phases by 

\beq
\phi_b~=~\phi_{orb}~-~\phi_{sh}~,  
\eeq

\noindent
and at the superhump maximum when $\phi_{sh}=0$ we simply have: $\phi_b=\phi_{orb}$.

\section {Superhump Amplitudes in Dwarf Novae at their Superoubtursts } 

\subsection {Data and their Analysis}

A search through the literature resulted in a sample of 26 dwarf novae 
(Table 1) with data on their superhump amplitudes suitable for our discussion. 
It includes 11 eclipsing systems with well determined orbital inclinations 
($i\geq 79^\circ$) taken from the Catalogue by Ritter and Kolb (2003) 
and 15 non-eclipsing systems with unknown, much lower inclinations. 

\begin{table}[h!]
{\parskip=0truept
\baselineskip=0pt {
\medskip
\centerline{Table 1}
\medskip
\centerline{ Amplitudes of Superhumps }
\smallskip
\centerline {Dwarf Novae at Superoutbursts }
\medskip
$$\offinterlineskip \tabskip=0pt
\vbox {\halign {\strut
\vrule width 0.5truemm #&	
\enskip\hfil#\enskip&	        
\vrule#&			
\enskip\hfil#\hfil\enskip&      
\vrule#&			
\enskip#\hfil\enskip&      
\vrule#&			
\enskip#\hfil\enskip&      
\vrule#&			
\enskip#\hfil\enskip&           
\vrule width 0.5 truemm # \cr	
\noalign {\hrule height 0.5truemm}
&&&&&&&&&&\cr
&Star\hfil&&$i$&&\hfil${\rm A_\circ~(mag)}$&&\hfil${\rm dA/dt~(mag/d)}$&&Data&\cr
&&&&&&&&&&\cr
\noalign {\hrule height 0.5truemm}
&&&&&&&&&&\cr
&  V1141 Aql &&      && $0.33\pm 0.02 $ && $-0.034\pm 0.006 $ && 1 &\cr
&   V877 Ara &&      && $0.26\pm 0.01 $ && $-0.019\pm 0.002 $ && 2 &\cr
&     TT Boo &&      && $0.24\pm 0.02 $ && $-0.008\pm 0.002 $ && 3 &\cr
&     OY Car && 83.3 && $0.51\pm 0.03 $ && $-0.041\pm 0.004 $ && 4,5 &\cr
&   V485 Cen &&      && $0.22\pm 0.01 $ && $-0.021\pm 0.002 $ && 6 &\cr
&      Z Cha && 80.2 && $0.49\pm 0.03 $ && $-0.046\pm 0.006 $ && 7 &\cr
&     EG Cnc &&      && $0.22\pm 0.02 $ && $-0.012\pm 0.002 $ && 8 &\cr
&   V503 Cyg &&      && $0.17\pm 0.01 $ && $-0.005\pm 0.003 $ && 9 &\cr
&     IX Dra &&      && $0.16\pm 0.03 $ && $-0.014\pm 0.007 $ && 10 &\cr
&     XZ Eri && 80.2 && $0.29\pm 0.05 $ && $-0.008\pm 0.008 $ && 11,12 &\cr
&   V660 Her &&      && $0.32\pm 0.02 $ && $-0.020\pm 0.003 $ && 13 &\cr
&     VW Hyi &&      && $0.30\pm 0.01 $ && $-0.019\pm 0.002 $ && 14 &\cr
&   V419 Lyr &&      && $0.30\pm 0.01 $ && $-0.015\pm 0.002 $ && 15 &\cr
&  V1159 Ori &&      && $0.20\pm 0.03 $ && $-0.030\pm 0.008 $ && 16 &\cr
&  V2051 Ori && 83.0 && $0.50\pm 0.10 $ &&	              && 17 &\cr
&     SW UMa &&      && $0.25\pm 0.01 $ && $-0.023\pm 0.002 $ && 18 &\cr
&     CY UMa &&      && $0.28\pm 0.02 $ && $-0.021\pm 0.005 $ && 19 &\cr
&     DI UMa &&      && $0.21\pm 0.02 $ && $-0.019\pm 0.003 $ && 20 &\cr
&     DV UMa && 84.0 && $0.61\pm 0.04 $ && $-0.032\pm 0.004 $ && 21 &\cr
&     IY UMa && 86.8 && $0.69\pm 0.08 $ && $-0.036\pm 0.008 $ && 22 &\cr
&     KS UMa &&      && $0.21\pm 0.03 $ && $-0.009\pm 0.005 $ && 23 &\cr
&     OU Vir && 79.2 && $0.34\pm 0.10 $ &&                    && 24 &\cr
& J1227+5139 && 83.9 && $0.51\pm 0.04 $ && $-0.028\pm 0.006 $ && 25 &\cr
& J1502+3334 && 88.9 && $0.53\pm 0.04 $ && $-0.018\pm 0.007 $ && 26 &\cr
& J1524+2209 && 82.8 && $0.30\pm 0.10 $ &&                    && 27 &\cr
& J1702+3229 && 82.4 && $0.60\pm 0.10 $ &&                    && 28,29 &\cr
&&&&&&&&&&\cr
\noalign {\hrule height 0.5truemm}
}}$$
Data sources: (1) Olech (2003) Fig.3. (2) Kato et al. (2003) Fig.3. 
(3) Olech et al. (2004b) Fig.5. (4) Krzemi{\'n}ski and Vogt (1985) Fig.2b. 
(5) Schoembs(1986) Fig.2a. (6) Olech (1997) Figs 2 and 4. 
(7) Warner and O'Donoghue (1988) Table 3 and Fig.5. 
(8) Patterson et al. (1998) Table 1. (9) Harvey et al. (1995) Fig.7. 
(10) Olech et al. (2004a) Fig.10. (11) Uemura et al. (2004) Fig.4. 
(12) Patterson et al. (2005) Fig.2. (13) Olech et al. (2005) Fig.3. 
-- continued on next page -- 
}}
\end{table}

Using earlier evidence (see Introduction) we assume that variations of the superhump 
amplitude with time during superoutburst can be represented by 

\beq
A~=~A_\circ~+~{{dA}\over{dt}}~\Delta t~,
\eeq

\noindent 
where $dA/dt<0$ and $\Delta t=0$ corresponds to the maximum amplitude $A_\circ$. 

The data on superhump amplitudes for objects listed in Table 1 were analysed 
using Eq.(7) and the resulting values of $A_\circ$ and $dA/dt$ are listed 
in columns 3 and 4 of this table. As could be expected they are tightly correlated 

\beq
{{dA}\over{dt}}~=~C~A_\circ~.  
\eeq 

\noindent 
where $C=0.063\pm 0.005$. 

At this point we can comment on the uncertainty of the values of $A_\circ$ 
resulting from the fact that the moments $\Delta t=0$ are 
not always precisely known. An uncertainty of $\pm 1$ day in the zero point 
of $\Delta t$ translates into an additional uncertainty of $A_\circ$ as 
$\pm dA/dt=\pm 0.063 A_\circ$ which is comparable or larger than formal errors 
listed in Table 1. Fortunately this additional uncertainty does not 
affect our main results. 

The procedure described above could not be applied to four high inclination 
systems (V2051 Ori, OU Vir, J1524+2209, and J1702+3229) for which only the 
{\it mean} light curves  were available. 
Their maximum superhump amplitudes $A_\circ$ were estimated from the 
mean amplitudes $<A>$ corresponding to the mean $<\Delta t>$ 

\noindent
Table 1. Data sources -- continued: 
(14) Haefner et al. (1979) Fig.7. (15) Rutkowski et al. (2007) Fig.4. 
(16) Patterson et al. (1995) Figs 4 and 5.
(17) Kiyota and Kato (1998) Fig.2. (18) Soejima et al. (2009) Figs 2,9,14. 
(19) Harvey and Patterson (1995) Fig.2. 
(20) Rutkowski et al. (2009) Figs 4 and 7.
(21) Patterson et al. (2000b) Fig.10. 
(22) Patterson et al. (2000a) Fig.2. 
(23) Olech et al. (2003) Fig.8. 
(24) Patterson et al. (2005) Fig.1. 
(25) Shears et al. (2008) Fig.3.
(26) Shears et al. (2010) Fig.3. (27) Kato et al. (2009) Fig.191.
(28) Boyd et al. (2006) Fig.8. (29) Kato et al. (2009) Fig.193. 
\medskip
\hrule

\noindent
using the following formula obtained by combining Eqs.(7) and (8)  

\beq
A_\circ~=~{{<A>}\over {1~-C<\Delta t>}}~.
\eeq

\begin{figure}[htb]
\epsfysize=7.5cm 
\hspace{1.0cm}
\epsfbox{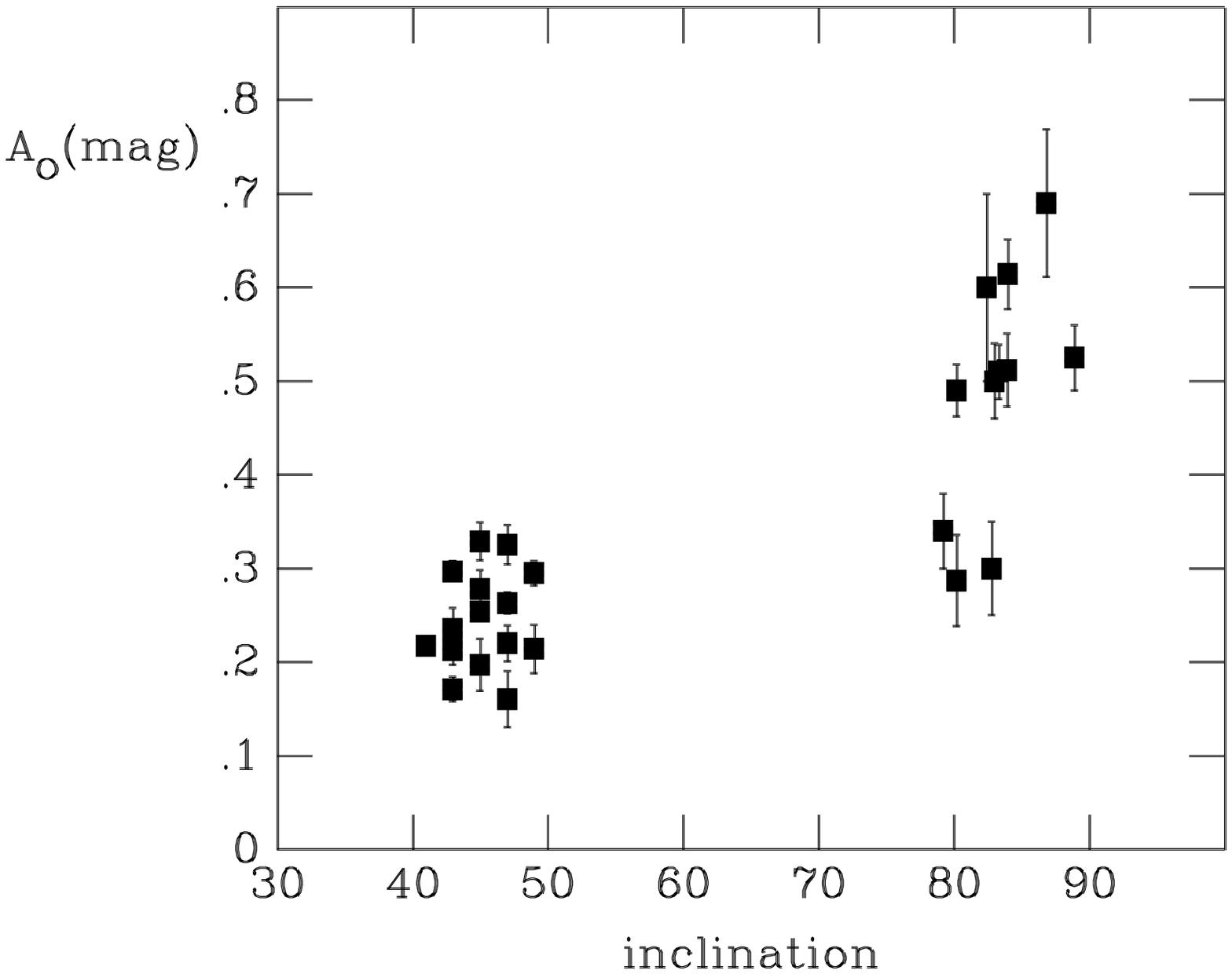} 
\vskip 5truemm
\FigCap { Maximum superhump amplitudes a a function of the orbital inclination.  
 Non-eclipsing systems with unknown inclinations are plotted between 
 $i=40^\circ$ and  $50^\circ$. }
\end{figure}

\subsection {Maximum Superhump Amplitudes versus Orbital Inclination} 

Fig.1 shows the maximum superhump amplitudes $A_\circ$ plotted against the 
orbital inclination. As we can see the values of $A_\circ$ for low inclination 
system cluster around $A_\circ \approx 0.25$. Those for high inclination, 
eclipsing systems are -- roughly -- 2 times larger, show more scatter 
and appear to increase from $A_\circ \approx 0.3$ at $i\approx 80^\circ$ 
to about $A_\circ \sim 0.6$ at the highest inclinations. 

This imposes important constraints on superhumps models and,  
in particular, eliminates all models which locate the superhump light source  
within the disk. 
As discussed above (Section 2) the superhump amplitude is defined 
with respect to the luminosity of the disk which depends strongly on inclination. 
Should the superhump source be located within the disk its luminosity would 
depend on inclination in the same way as the luminosity of the disk and the 
observed superhump amplitude would be independent of inclination. 
 
Using Eq.(5) we now normalize the maximum superhump amplitudes $A_\circ$ 
to the mean luminosity of the disk. For non-eclipsing systems, with inclinations 
which are generally lower than about $70^\circ$, assuming their random 
distribution we adopt: $<i>=<i~\sin i>/<\sin i>\approx 45^\circ$.  

The resulting normalized amplitudes $A_n$ are plotted against orbital inclination 
in Fig.2. Compared to Fig.1 the situation is now different. 
The superhump amplitudes of low inclination systems cluster around 
$<A_n>=0.34\pm 0.02$ with {\it r.m.s.} dispersion of individual values 
$\sigma =\pm 0.07$. This value of $A_n$ can then 
be considered as representative for superhumps observed during superoutbursts. 
The amplitudes of high inclination systems cluster around 
$<A_n>=0.17\pm 0.01$, show scatter $\sigma =\pm 0.04$ which is much smaller 
than that in Fig.1, and do not show any dependence on inclination. 
The obvious interpretation is that the superhump source is located close 
to the disk surface but extends sufficiently high above that surface 
so that in systems with high inclinations it is {\it partly} obscured 
by the disk edge. 

\begin{figure}[htb]
\epsfysize=5.0cm 
\hspace{1.5cm}
\epsfbox{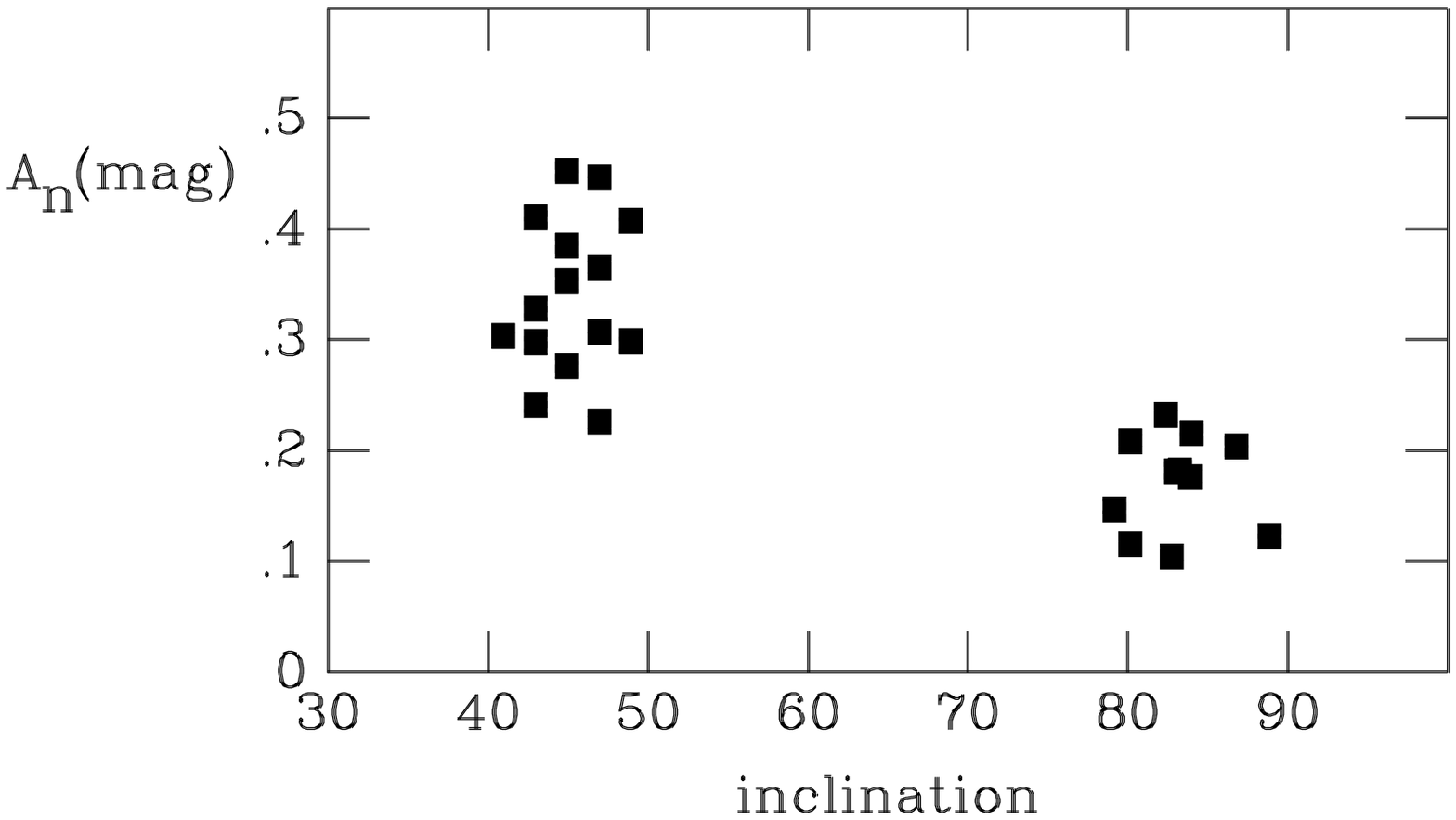} 
\vskip 5truemm
\FigCap { Maximum superhump amplitudes normalized to the average disk luminosity 
 as a function of the orbital inclination. } 
\end{figure}

Not included in our discussion, so far, was U Gem. During its 1985 superoutburst  
it showed superhumps with amplitude $A\approx 0.3$ mag. (Smak and Waagen 2004).  
With $i=69^\circ$ this places it in Fig.1 just between the low and high 
inclinations systems. Unlike in all other dwarf novae, however, the superhump 
amplitude of U Gem stayed constant throughout the entire superoutburst.  
There is no obvious explanation of this peculiar behavior. 

\subsection {Dependence of Superhump Amplitudes on Beat Phase}  

The superhump amplitude can be modulated with beat phase due to various effects. 
Two such effects, suggested by earlier evidence, are considered here. 

It was shown earlier (Smak 2009b) that in systems with the highest orbital 
inclinations ($i>82^\circ$) their disk luminosity during superoutbursts 
is modulated with the beat period. This was intepreted as being due to 
a non-axisymmetric structure of the disk, involving the azimuthal dependence 
of the vertical thickness of its outer edge. 
At that time there were only three systems (OY Car, DV UMa and IY UMa) showing 
such modulation. Now we can strengthen our earlier conclusion by adding another 
deeply eclipsing system: J1227+5139. Using data from Shears et al. (2008) 
and reducing them in the same way as before (Smak 2009b, Section 2) we obtain 
the residual magnitudes which show clear dependence on the beat phase (Fig.3). 
The two parameters describing this dependence: the half-amplitude 
$A({\rm mag})=0.18\pm 0.03$ and $\phi_b^{max}=0.68\pm 0.03$ are -- within errors -- 
identical with those obtained earlier for three other systems. Using now all four systems 
we get: $<A({\rm mag})>=0.18\pm 0.01$ and $<\phi_b^{max}>=0.65\pm 0.02$. 
An obvious question is: do the amplitudes of superhumps show similar effect? 

\begin{figure}[htb]
\epsfysize=4.0cm 
\hspace{2.0cm}
\epsfbox{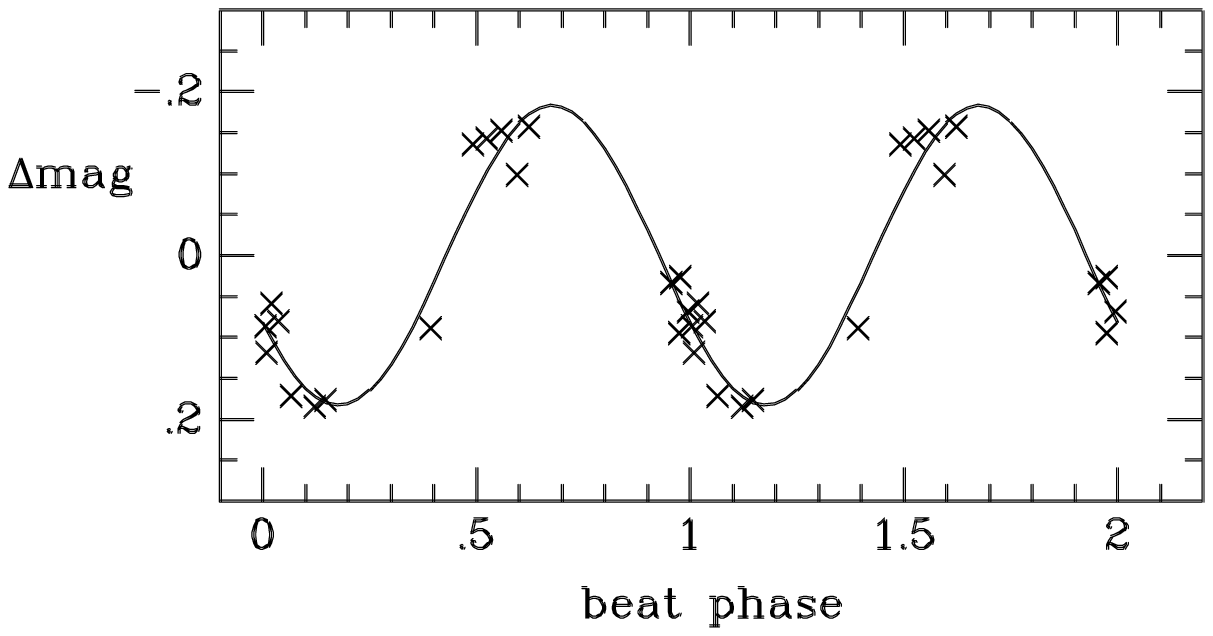} 
\vskip 5truemm
\FigCap { Residual disk magnitudes for J1227+5139 are plotted against 
 the beat phase. Solid line is the cosine curve with parameters given in the text. }
\end{figure}

Another effect, to be considered here, is suggested by earlier results 
concerning superhumps observed in U Gem during its December 1985 superoutburst. 
It was found (Smak 2006) that their amplitude was modulated with beat phase 
in the form of a {\it double} cosine wave. 
An obvious question is: do other dwarf novae show similar modulation?

To answer those questions we analyze data on superhumps for six high inclination 
and -- for comparison -- for four low inclination systems with sufficient 
coverage in beat phases. 
The superhump amplitudes observed at $\Delta t$ and at beat phase $\phi_b$ 
are represented with 

\beq
A~=~A_\circ~+~{{dA}\over{dt}}~\Delta t~
        +~A_1~\cos ~\bigl(\phi_b~-~\phi_{b,1}^{max}\bigr)~
        +~A_2~\cos ~\Bigl[~2~\bigl(\phi_b~-~\phi_{b,2}^{max}\bigr)~\Bigr]~.  
\eeq

\noindent
The last term in this equation describes the "double-$\phi_b$" modulation 
with two maxima at $\phi_{b,2}^{max}$ and $\phi_{b,2}^{max}+0.5$ and 
two minima at $\phi_{b,2}^{max}\pm 0.25$. 

The residuals $\Delta A$, representing the "single-$\phi_b$" and 
"double-$\phi_b$" modulations, are calculated as 

\beq
\Delta A~=~A~-\left(~A_\circ~+~{{dA}\over{dt}}~\Delta t~\right )~. 
\eeq

\begin{figure}[htb]
\epsfysize=12.0cm 
\hspace{2.0cm}
\epsfbox{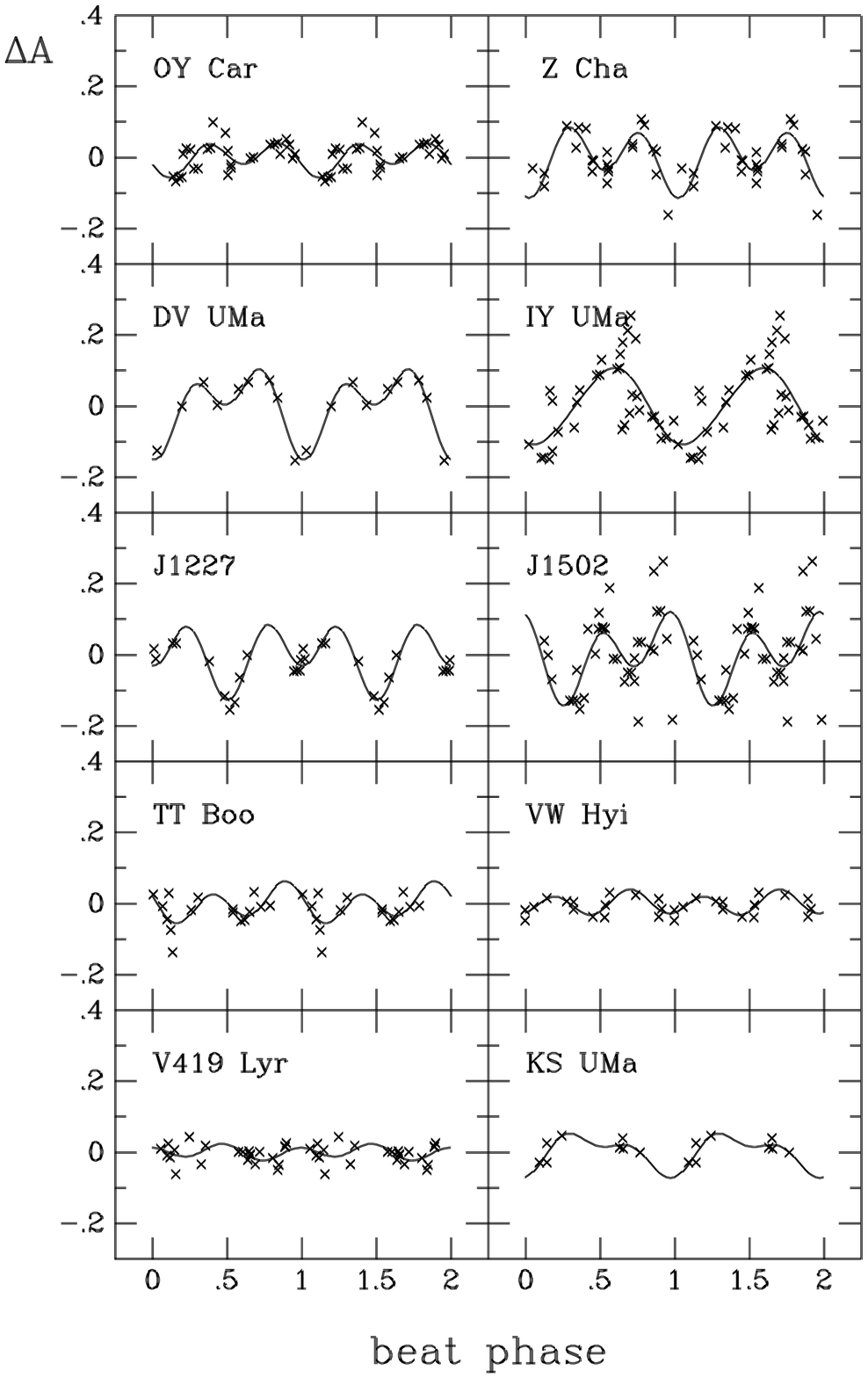} 
\vskip 5truemm
\FigCap { Residual superhump amplitudes vs. beat phase 
 for 6 high inclination and 4 low inclination systems. 
 Zero points of $\phi_b$ for TT Boo, VW Hyi, V419 Lyr and KS UMa are arbitrary. 
 Solid lines are cosine and double cosine curves with parameters 
 obtained from solutions using Eq.(10) and listed in Table 2. 
 See text for details. } 
\end{figure}

\begin{table}[h!]
{\parskip=0truept
\baselineskip=0pt {
\medskip
\centerline{Table 2}
\medskip
\centerline{ Amplitudes of Superhumps }
\smallskip
\centerline { Modulation with Beat Phase }
\medskip
$$\offinterlineskip \tabskip=0pt
\vbox {\halign {\strut
\vrule width 0.5truemm #&	
\enskip\hfil#\enskip&	        
\vrule#&			
\enskip#\enskip&      
\vrule#&			
\enskip#\enskip&      
\vrule#&			
\enskip#\enskip&      
\vrule#&			
\enskip#\enskip&           
\vrule width 0.5 truemm # \cr	
\noalign {\hrule height 0.5truemm}
&&&&&&&&&&\cr
&Star\hfil&&\hfil$A_1$\hfil&&\hfil$\phi_{b,1}^{max}$\hfil&&\hfil$A_2$\hfil&&\hfil$\phi_{b,2}^{max}$\hfil&\cr
&&&&&&&&&&\cr
\noalign {\hrule height 0.5truemm}
&&&&&&&&&&\cr
&    OY Car && $0.02\pm 0.01$ && $0.63\pm 0.13$ && $0.04\pm 0.01$ && $0.36\pm 0.12$ &\cr
&     Z Cha && $0.04\pm 0.01$ && $0.50\pm 0.05$ && $0.07\pm 0.03$ && $0.28\pm 0.03$ &\cr
&    DV UMa && $0.08\pm 0.02$ && $0.55\pm 0.03$ && $0.07\pm 0.01$ && $0.26\pm 0.02$ &\cr
&    IY UMa && $0.11\pm 0.03$ && $0.58\pm 0.05$ && $0.01\pm 0.03$ && $0.20\pm 0.11$ &\cr
&J1227+5139 && $0.05\pm 0.02$ && $0.99\pm 0.06$ && $0.08\pm 0.03$ && $0.25\pm 0.03$ &\cr
&J1502+3334 && $0.06\pm 0.03$ && $0.82\pm 0.16$ && $0.09\pm 0.03$ && $0.50\pm 0.02$ &\cr
&    TT Boo && $0.02\pm 0.03$ && \hfil	    && $0.04\pm 0.03$ && \hfil	    &\cr
&    VW Hyi && $0.01\pm 0.01$ && \hfil	    && $0.03\pm 0.01$ && \hfil	    &\cr
&  V419 Lyr && $0.01\pm 0.01$ && \hfil	    && $0.02\pm 0.03$ && \hfil	    &\cr
&    KS UMa && $0.05\pm 0.05$ && \hfil	    && $0.03\pm 0.08$ && \hfil	    &\cr
&&&&&&&&&&\cr
\noalign {\hrule height 0.5truemm}
}}$$
}}
\end{table}

Results are listed in Table 2 and shown in Fig.4. 
Before discussing them we must make few comments. In the case of low inclination 
systems with no orbital elements the zero point of $\phi_{orb}$ and, consequently, 
of $\phi_b$ are arbitrary. The orbital period of VW Hyi was taken from Vogt (1974), 
while for three other systems it was calculated from $P_{sh}$ using formula 
given by Menninckent et al. (1999). 

The expected "single-$\phi_b$" modulation is present in all high inclination 
systems (although in the case of OY Car it is barely significant).  
The mean value of $A_1$ for those six systems is: $<A_1>=0.06$ (with {\it r.m.s.} 
dispersion of individual values being $\sigma=0.03$), while in the case of the 
four low inclination systems it is only $<A_1>=0.02$ (with $\sigma=0.02$). 
The mean value of $\phi_{b,1}^{max}$ for 5 high inclination systems -- excluding 
the deviating value for J1227+5139 -- is $<\phi_{b,1}^{max}>=0.62$ (with 
$\sigma =0.06$) or -- including J1227+5139 -- $<\phi_{b,1}^{max}>=0.68$ 
(with $\sigma =0.08$). 
In either case this is consistent with $<\phi_b^{max}>=0.65\pm 0.02$ obtained 
from disk luminosity modulation (see above). 

Turning to the "double-$\phi_b$" modulation we find that it is present in five 
high inclination systems, the only exception being IY UMa.  
The mean value of $A_2$ -- including IY UMa -- is $<A_2>=0.06$ ($\sigma=0.03$) 
to be compared with only $<A_2>=0.03$ ($\sigma=0.02$) for low inclinations systems. 
The mean value of $\phi_{b,2}^{max}$ -- excluding IY UMa -- 
is $<\phi_{b,2}^{max}>=0.33$ (with $\sigma =0.09$) or -- excluding also 
the deviating value for J1502+3334 -- $<\phi_{b,2}^{max}>=0.29$ 
(with $\sigma =0.04$). We adopt: $<\phi_{b,2}^{max}>=0.30$. 
It may be added that the value obtained earlier for U Gem (Smak 2006), namely   
$\phi_{b,2}^{max}\approx 0.15\pm 0.13$ is -- within errors -- practically 
the same. 

\begin{figure}[htb]
\epsfysize=5.0cm 
\hspace{3.0cm}
\epsfbox{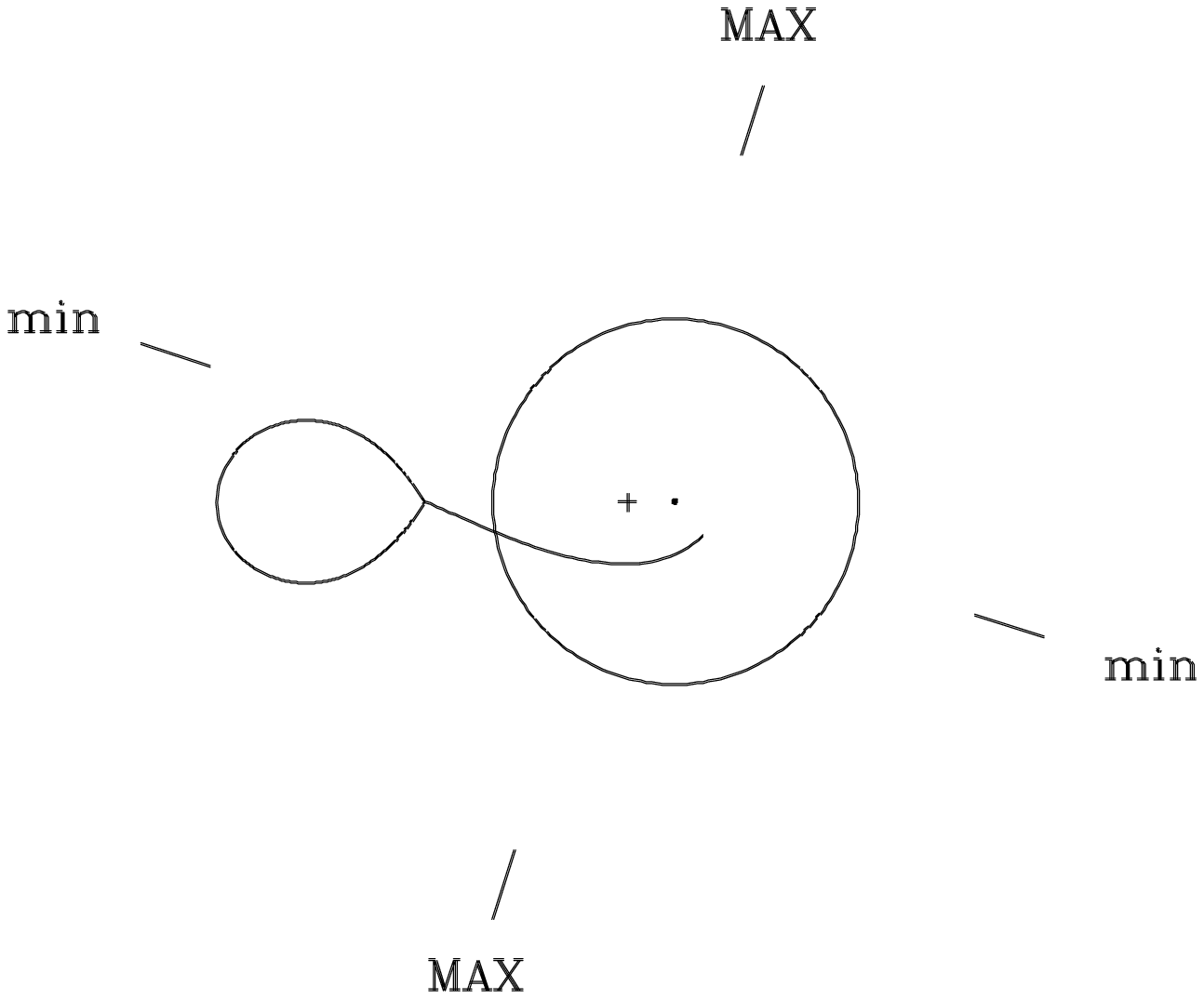} 
\vskip 5truemm
\FigCap { Schematic model of a cataclysmic binary with $q=0.15$  
showing the secondary, the disk (for simplicity shown circular), and 
the stream trajectory. The position of the center of mass is marked with a cross. 
Phases of highest and lowest superhump amplitudes are marked with 
"MAX" and "min". } 
\end{figure}

The presence of the "double-$\phi_b$" modulation and, in particular, the value 
of $<\phi_{b,2}^{max}>$ provide an important clue concerning the location 
of the superhump light source. 
Fig.5 presents schematic view of a system with $q=0.15$ (typical for 
systems considered here). Shown there are the two phases of the highest 
superhump amplitudes: $\phi_b^{max}=\phi_{orb}^{max}=0.30$ and 0.80
and two other phases corresponding to the lowest amplitudes: 
$\phi_b^{min}=\phi_{orb}^{min}=0.05$ and 0.55.   
The conclusion is obvious: the orientation of the superhump light source 
is correlated with the direction of the stream. More specifically: 
the maximum superhump amplitude occurs when the line of sight is perpendicular 
to the stream (including its part overflowing the disk), while the minimum 
amplitude -- when the line of sight is parallel to the stream.

\section {Superhump Amplitudes in Permanent Superhumpers }

Table 3 contains representative sample of ten permanent superhumpers 
including data on their superhump amplitudes. 
Regretfully, there are only two eclipsing systems with moderately high 
inclination. Therefore not much weight can be given to the fact that their 
amplitudes are practically the same as at lower inclinations. 

\begin{table}[h!]
{\parskip=0truept
\baselineskip=0pt {
\medskip
\centerline{Table 3}
\medskip
\centerline{ Amplitudes of Superhumps }
\smallskip
\centerline {Nova-Like Permanent Superhumpers }
\medskip
$$\offinterlineskip \tabskip=0pt
\vbox {\halign {\strut
\vrule width 0.5truemm #&	
\quad\hfil#\quad&	        
\vrule#&			
\enskip\hfil#\hfil\enskip&      
\vrule#&			
\quad\hfil#\hfil\enskip&        
\vrule#&		        
\quad#\hfil\enskip&             
\vrule width 0.5 truemm # \cr	
\noalign {\hrule height 0.5truemm}
&&&&&&&&\cr
&Star\hfil&& $i$ && A(mag) && \hfil Data source &\cr
&&&&&&&&\cr
\noalign {\hrule height 0.5truemm}
&&&&&&&&\cr
&V603 Aql && 13   && 0.12 && Patterson et al. (1993) Table 1 &\cr
&  TT Ari &&      && 0.15 && Skillman et al. (1998) Fig.2 &\cr
&V592 Cas &&      && 0.15 && Taylor et al. (1998) Fig.1 &\cr
&  TV Col &&      && 0.08 && Retter et al. (2003) Fig. 4 &\cr
&         &&      &&      && Hellier (1993) &\cr
&  BB Dor &&      && 0.08 && Patterson et al. (2005) Fig.7 &\cr
&V795 Her &&      && 0.22 && Patterson and Skillman (1994) &\cr
&  BK Lyn &&      && 0.06 && Skillman and Patterson (1993) Fig.6 &\cr
&  MV Lyr && 12   && 0.11 && Borisov (1992) Fig.4 &\cr
&V348 Pup && 81.1 && 0.11 && Rolfe et al.(2000) Fig.6  &\cr
&  DW UMa && 82.0 && 0.13 && Stanishev et al. (2004) Fig.4  &\cr
&         &&              &&      && Patterson et al. (2002) Fig.15  &\cr
&&&&&&&&\cr
\noalign {\hrule height 0.5truemm}
}}$$
}}
\end{table}

The mean superhump amplitude for objects listed in Table 3 is $<A>=0.12$ mag.  
This is much lower than the maximum amplitudes observed during superoutbursts, 
comparable to the amplitudes observed near the end of a superoutburst. 
Two possible explanations of this difference could be proposed. The lower 
superhump amplitudes in permanent superhumpers could be due to a higher 
mass transfer rate and higher luminosity of the disk. On the other hand, 
it could be speculated that the higher maximum superhump amplitudes in dwarf 
novae are due to a sudden enhancement of the mass transfer rate at the begining 
of a superoutburst.

\section { The Nature of Superhumps }  

\subsection { The Tidal-Resonance Model }  

The commonly accepted tidal-resonance model, first proposed 
by Whitehurst (1988) and Hirose and Osaki (1990), explains superhumps as being 
due to tidal effects in the outer parts of accretion disks leading 
-- via the 3:1 resonance -- to the formation of an eccentric outer ring 
undergoing apsidal motion. 
This model and, in particular, the results of numerous 2D and 3D SPH simulations 
(cf. Smith et al. 2007 and references therein) reproduce the observed 
superhump periods and correlations of the superhump period excess with the orbital 
period and the mass-ratio. 
This suggests that the basic "clock" which defines the superhump periods  
is probably provided by the apsidal motion (but see point (1) below). 

On the other hand, however, the tidal-resonance model fails to explain  
all other important facts:  

{\parskip=0truept {
(1) The 3:1 resonance, which is the crucial ingredient of the tidal-resonance 
model, can occur only in systems with mass ratios $q=M_2/M_1$ smaller 
than $q_{crit}=0.25$ (cf. Whitehurst 1988, Osaki 2005 and references therein; 
see also Smak 2006 for a critical review of attempts to increase $q_{crit}$ 
up to 0.33). 
The condition $q<q_{crit}$ is fulfilled by the short period dwarf novae of the 
SU UMa type. There are, however, other systems with mass ratios higher than 
$q_{crit}$ which also show superhumps. 
Examples: dwarf novae U Gem with $q=0.36\pm 0.02$ (Smak 2001), 
and TU Men with $q\approx 0.5\pm 0.2$ or $q>0.41\pm 0.08$ (Smak 2006), 
and the growing number of permanent superhumpers with longer orbital periods 
indicating much higher mass ratios; among them: 
MV Lyr with $q=0.43_{-0.13}^{+0.19}$ (Skillman et al. 1995), 
DW UMa with $q=0.39\pm 0.12$ (Araujo-Betancor et al. 2003), 
BH Lyn with $q=0.45_{-0.10}^{+0.15}$ (Hoard and Szkody 1997), and 
TV Col with $q=0.62-0.93$ (Hellier 1993) or $q=0.92\pm 0.12$ (Retter et al. 2003). 

(2) Numerical 2D and 3D SPH simulations produce "superhumps" with visual amplitudes 
which are smaller than $A=0.10$ (from 2D models), or as small as $A=0.03-0.04$ 
(from 3D models) (see Smak 2009a and references therein).  
They are by factor $4-10$ smaller than $<A_n>=0.34\pm 0.02$ obtained 
from observations in Section 3.2.  

(3) The dependence of superhump amplitudes on the orbital inclination 
(Section 3.2) eliminates all models which locate the source of superhumps 
within the disk. 

(4) The presence and characteristics of obscuration effects which affect 
superhump amplitudes in systems with high orbital inclinations 
(Sections 3.2 and 3.3) imply also that the superhump source is not located 
within the disk but extends sufficiently high above its surface to avoid 
full obscuration. 

(5) The presence of the "double-$\phi_b$" modulation of superhump amplitudes 
related to the orientation of the stream (Section 3.3) finds no explanation 
within the tidal-resonance model. 
}}
\parskip=12truept 

\subsection { An Alternative Interpretation of Superhumps }

Using evidence available earlier it was proposed (Smak 2009b) that  
superhumps are due to periodically modulated dissipation of the kinetic energy 
of the stream, the essential ingredients of this interpretation being as follows: 

{\parskip=0truept {
(1) The outer parts of the disk have a non-axisymmetric structure, involving 
the azimuthal dependence of their vertical thickness, rotating -- in the inertial 
frame -- with $P_b$. 

(2) This causes the irradiation of the secondary component to be modulated 
with $P_{irr}$ (related to $P_b$ and $P_{orb}$), resulting in 
periodic variations of the mass transfer rate with $P_{sh}\approx P_{irr}$. 

(3) Around superhump maximum the stream overflows the surface of the disk 
and -- unlike in the case of the "standard" hot spot -- its energy is dissipated 
along its trajectory above (and below) the disk. 

(4) The periodically variable dissipation of the kinetic energy of the stream  
is observed in the form of superhumps. 
}}
\parskip=12truept 

Results presented in this paper provide further arguments in favor of this 
interpretation. In particular: 

{\parskip=0truept {
(1) The presence and characteristics of obscuration effects which affect 
superhump amplitudes in systems with high orbital inclinations 
(Sections 3.2 and 3.3) imply that the superhump light source is indeed located 
above the surface of the disk. 

(2) The "double-$\phi_b$" modulation of the superhump amplitudes (Section 3.3) 
implies that the orientation of the superhump light source is -- as could be 
expected -- correlated with the direction of the stream. 
In particular, when the line of sight is perpendicular to the stream trajectory, 
the superhump must be brighter due to larger projected surface area of the 
overflowing parts of the stream. 
}}
\parskip=12truept

\section {Conclusions } 

The new results concerning superhump amplitudes, their dependence on orbital 
inclination, and their modulation with the beat phase, presented in this paper,  
provide further arguments against the commonly accepted tidal-resonance model 
for superhumps. 
At the same time they support the alternative interpretation of superhumps 
involving the periodically modulated dissipation of the kinetic energy of 
the stream. 

\bigskip

\begin {references} 

\refitem {Araujo-Betancor, S. et al.} {2003} {\ApJ} {583} {437} 
\refitem {Borisov, G.V.} {1992} {\AA} {261} {154} 
\refitem {Boyd, D., Oksanen, A., Henden, A.} {2006} {\it {J.Br.Astron.Assoc.}} 
          {116} {187}
\refitem {Haefner, R., Schoembs, R., Vogt, N.} {1979} {\AA} {77} {7}
\refitem {Harvey, D.A., Patterson, J.} {1995} {\PASP} {107} {1055}
\refitem {Harvey, D.A., Skillman, D.R., Patterson, J., Ringwald, F.A.} 
          {1995} {\PASP} {107} {551}
\refitem {Hellier, C.} {1993} {\MNRAS} {264} {132}
\refitem {Hellier, C.} {2001} {{\it Cataclysmic Variable Stars} {\rm (Springer)} } {~} {~} 
\refitem {Hirose, M., Osaki, Y.} {1990} {\PASJ} {42} {135} 
\refitem {Hoard, D.W., Szkody, P.} {1997} {\ApJ} {481} {433}
\refitem {Kato, T. et. al.} {2003} {\MNRAS} {339} {861}
\refitem {Kato, T. et. al.} {2009} {\PASJ} {61} {S395}
\refitem {Kiyota, S., Kato, T.} {1998} {\it Inf.Bull.Var.Stars} {~} {No.4664}
\refitem {Krzemi{\'n}ski, W., Vogt, N.} {1985} {\AA} {144} {124}
\refitem {Menninckent, R.E., Matsumoto, K., Arenas, J.} {1999} {\AA} {348} {466} 
\refitem {Olech, A.} {1997} {\Acta} {47} {281} 
\refitem {Olech, A.} {2003} {\Acta} {53} {85} 
\refitem {Olech, A., Schwarzenberg-Czerny, A., K{\c e}dzierski, P.,  
          Z{\l}oczewski, K.,Mularczyk, K., Wi{\'s}niewski, M.} {2003} {\Acta} 
          {53} {175} 
\refitem {Olech, A., Z{\l}oczewski, K., Mularczyk, K.,K{\c e}dzierski, P., 
          Wi{\'s}niewski, M., Stachowski, G.} {2004a} {\Acta} {54} {57} 
\refitem {Olech, A., Cook, L.M., Z{\l}oczewski, K., Mularczyk, K., 
          K{\c e}dzierski, P., Udalski, A., Wi{\'s}niewski, M.} {2004b} 
          {\Acta} {54} {233} 
\refitem {Olech, A., Z{\l}oczewski, K., Cook, L.M., Mularczyk, K., 
          K{\c e}dzierski, P., Wi{\'s}niewski, M.} {2005} {\Acta} {55} {237} 
\refitem {Osaki, Y.} {2005} {{\it Proc.Japan Academy, Ser.B}} {81} {291}
\refitem {Patterson, J.} {1999} { {\it Disk Instabilities in Close Binary 
          Systems}, {\rm eds. S.Mineshige and J.C.Wheeler (Tokyo: Universal
          Academy Press)} } {~} {61}
\refitem {Patterson, J., Thomas, G., Skillman, D.R., Diaz, M.} {1993} 
          {\ApJS} {86} {235}
\refitem {Patterson, J., Skillman, D.R.} {1994} {\PASP} {106} {1141}
\refitem {Patterson, J., Jablonsky, F., Koen, C., O'Donoghue, D., Skillman, D.R.} 
          {1995} {\PASP} {107} {1183} 
\refitem {Patterson, J. et al.} {1998} {\PASP} {110} {1290}
\refitem {Patterson, J.,Kemp, J., Jensen, L., Vanmunster, T., Skillman, D.R.,
	  Martin, B., Fried, R., Thorstensen, J.R.} {2000a} {\PASP} {112} {1567}
\refitem {Patterson, J., Vanmunster, T., Skillman, D.R., Jensen, L., 
          Stull, J., Martin, B., Cook, L.M., Kemp, J., Knigge, C.} 
          {2000b} {\PASP} {112} {1584} 
\refitem {Patterson, J. et al.} {2002} {\PASP} {114} {1364}
\refitem {Patterson, J. et al.} {2005} {\PASP} {117} {1204} 
\refitem {Retter, A., Hellier, C., Augusteijn, T., Naylor, T., Bedding, T.R., 
          Bembrick, C., McCormick, J., Velthuis, F.} {2003} {\MNRAS} {340} {679} 
\refitem {Ritter, H., Kolb, U.} {2003} {\AA} {404} {301} (update RKcat. 7.14, 2010)
\refitem {Rolfe, D.J., Haswell, C.A., Patterson, J.} {2000} {\MNRAS} {317} {759} 
\refitem {Rutkowski, A., Olech, A., Mularczyk, K., Boyd, D., Koff, R.,
          Wi{\'s}niewski, M.} {2007} {\Acta} {57} {267}
\refitem {Rutkowski, A., Olech, A., Wi{\'s}niewski, M., Pietrukowicz, P.,
          Pala, J., Poleski, R.} {2009} {\AA} {497} {437}
\refitem {Schoembs, R.} {1986} {\AA} {158} {233} 
\refitem {Semeniuk, I.} {1980} {\AAS} {39} {29}
\refitem {Shears, J., Brady, S., Foote, J., Starkey, D., Vanmunster, T.} {2008}
          {\it {J.Br.Astron.Assoc.}} {118} {288}
\refitem {Shears, J. et al.} {2010} {{\rm arXiv}} {~} {1005.3219}
\refitem {Skillman, D.R., Patterson, J.} {1993} {\ApJ} {417} {298} 
\refitem {Skillman, D.R., Patterson, J., Thorstensen, J.R.} {1995} {\PASP} 
          {107} {545}
\refitem {Skillman, D.R. et al.} {1998} {\ApJ} {503} {L67}
\refitem {Smak, J.} {2001}  {\Acta} {51} {279}
\refitem {Smak, J.} {2006}  {\Acta} {56} {365}
\refitem {Smak, J.} {2009a} {\Acta} {59} {103} 
\refitem {Smak, J.} {2009b} {\Acta} {59} {121}
\refitem {Smak, J., Waagen, E.O.} {2004} {\Acta} {54} {433}
\refitem {Smith, A.J., Haswell, C.A., Murray, J.R., Truss, M.R., 
          Foulkes, S.B.} {2007} {\MNRAS} {378} {785}
\refitem {Soejima, Y. et al.} {2009} {\PASJ} {61} {659} 
\refitem {Stanishev, V., Kraicheva, Z., Boffin, H.M.J., Genkov, V., Papadaki, C., 
          Carpano, S.} {2004} {\AA} {416} {1057} 
\refitem {Still, M., Howell, S.B., Wood, M.A., Cannizzo, J.K., Smale, A.P.}
          {2010} {\ApJ} {717} {L113}
\refitem {Taylor, C.J., Thorstensen, J.R., Patterson, J., Fried, R.E.,  
          Vanmunster, T., Harvey, D.A., Skillman, D.R., Jensen, L., Shugarov, S.} 
          {1998} {\PASP} {110} {1148}
\refitem {Uemura, M. et al.} {2004} {\PASJ} {56} {141} 
\refitem {Vogt, N.} {1974} {\AA} {36} {369}
\refitem {Warner, B.} {1975} {\MNRAS} {170} {219} 
\refitem {Warner, B.} {1985} { {\it Interacting Binaries,} {\rm eds. P.P.Eggleton, 
  and J.E.Pringle (Reidel)} } {~} {367} 
\refitem {Warner, B.} {1995} {\it Cataclysmic Variable Stars} {~} {~} 
         (Cambridge University Press). 
\refitem {Warner, B., O'Donoghue, D.} {1988} {\MNRAS} {233} {705}
\refitem {Whitehurst, R.} {1988} {\MNRAS} {232} {35} 

\end {references}

\end{document}